\documentclass[conference]{IEEEtran}
\IEEEoverridecommandlockouts
\usepackage{cite}
\usepackage{amsmath,amssymb,amsfonts}
\usepackage{algorithmic}
\usepackage{textcomp}
\usepackage{xcolor}
\usepackage{graphicx}
\usepackage{multirow}
\usepackage{makecell}
\usepackage{csquotes}
\usepackage{framed}
\usepackage{array}
\usepackage[bookmarks=false]{hyperref}
\usepackage[capitalise]{cleveref}
\usepackage{dblfloatfix}

\newcolumntype{L}[1]{>{\raggedright\let\newline\\\arraybackslash\hspace{0pt}}m{#1}}
\newcolumntype{C}[1]{>{\centering\let\newline\\\arraybackslash\hspace{0pt}}m{#1}}
\newcolumntype{R}[1]{>{\raggedleft\let\newline\\\arraybackslash\hspace{0pt}}m{#1}}

\hypersetup{
    colorlinks=true,
    linkcolor=blue,
    urlcolor=blue
}

\makeatletter
\renewcommand{\@IEEEsectpunct}{\ \,}%
\makeatother

\def\BibTeX{{\rm B\kern-.05em{\sc i\kern-.025em b}\kern-.08em
    T\kern-.1667em\lower.7ex\hbox{E}\kern-.125emX}}

\begin{document}

\title{\huge DebtViz: A Tool for Identifying, Measuring, Visualizing, and Monitoring Self-Admitted Technical Debt
}
 
\author{Yikun Li, Mohamed Soliman, Paris Avgeriou, Maarten van Ittersum \\
\IEEEauthorblockA{Bernoulli Institute for Mathematics, Computer Science and Artificial Intelligence \\
University of Groningen \\
Groningen, The Netherlands \\
\{yikun.li, m.a.m.soliman, p.avgeriou, m.van.ittersum\}@rug.nl}
}

\maketitle

\begin{abstract}
Technical debt, specifically Self-Admitted Technical Debt (SATD), remains a significant challenge for software developers and managers due to its potential to adversely affect long-term software maintainability. 
Although various approaches exist to identify SATD, tools for its comprehensive management are notably lacking. 
This paper presents DebtViz, an innovative SATD tool designed to automatically detect, classify, visualize and monitor various types of SATD in source code comments and issue tracking systems.
DebtViz employs a Convolutional Neural Network-based approach for detection and a deconvolution technique for keyword extraction. 
The tool is structured into a back-end service for data collection and pre-processing, a SATD classifier for data categorization, and a front-end module for user interaction.
DebtViz not only makes the management of SATD more efficient but also provides in-depth insights into the state of SATD within software systems, fostering informed decision-making on managing it.
The scalability and deployability of DebtViz also make it a practical tool for both developers and managers in diverse software development environments.
The source code of DebtViz is available at \url{https://github.com/yikun-li/visdom-satd-management-system} and the demo of DebtViz is at \url{https://youtu.be/QXH6Bj0HQew}.
\end{abstract}

\begin{IEEEkeywords}
self-admitted technical debt, technical debt management, technical debt visualization 
\end{IEEEkeywords}

\section{Introduction}

Technical debt, a prevalent concept in software development, expresses the trade-offs often made between ideal development practices and short-term project needs \cite{avgeriou_et_al:DR:2016:6693}.
These trade-offs may involve hasty decisions, shortcuts, or less-than-ideal solutions geared toward expediting feature implementation or reducing development time.
While these quick-fixes may serve immediate needs, they can negatively impact long-term software maintenance.
In certain situations, consciously incurring technical debt can offer short-term advantages, provided it is managed effectively \cite{allman2012managing}. 
However, un-controlled accumulation of such debt often evolves into formidable maintenance challenges over time.
Thus, adept management of both intentional and unintentional technical debt is pivotal to maintain software quality and restrict the increasing cost of change \cite{lim2012balancing}.

\textit{Self-Admitted Technical Debt} (SATD) is a specific variant of technical debt, where developers voluntarily document technical debt within various software artifacts like source code comments, commit messages, issue tracking systems, or pull requests \cite{potdar2014exploratory, zampetti2021self}.
For example, a developer might mention pending tasks within a code comment like \textit{``we need to remove the dead code''}, or acknowledge a method's complexity with \textit{``this method is hard to understand and needs to be simplified''}.
Systematically detecting SATD helps to make such items explicit to developers and managers, assists in formulating plans for their resolution and thereby contributes to enhancement of maintainability and evolvability.

Existing research in this field, primarily focuses on identifying SATD using source code comments \cite{da2017using,ren2019neural}.
More recent studies also explore SATD detection from other software artifacts \cite{dai2017detecting,li2022identifying,li2022automatic}, such as issue tracking systems.
Despite these advances, 
practical tools to assist developers in detecting and managing different types of SATD are markedly absent. While research studies present several machine learning models, there is only one Eclipse plugin proposed for identifying and presenting SATD in code comments \cite{liu2018satd};
even this tool is constrained to handling SATD in code comments and lacks a comprehensive dashboard for system-wide SATD monitoring.
Thus, there is no comprehensive tool, capable of gathering, analyzing, visualizing and monitoring SATD from multiple software artifacts.

\begin{figure*}[t]
\centering
\includegraphics[width=\linewidth]{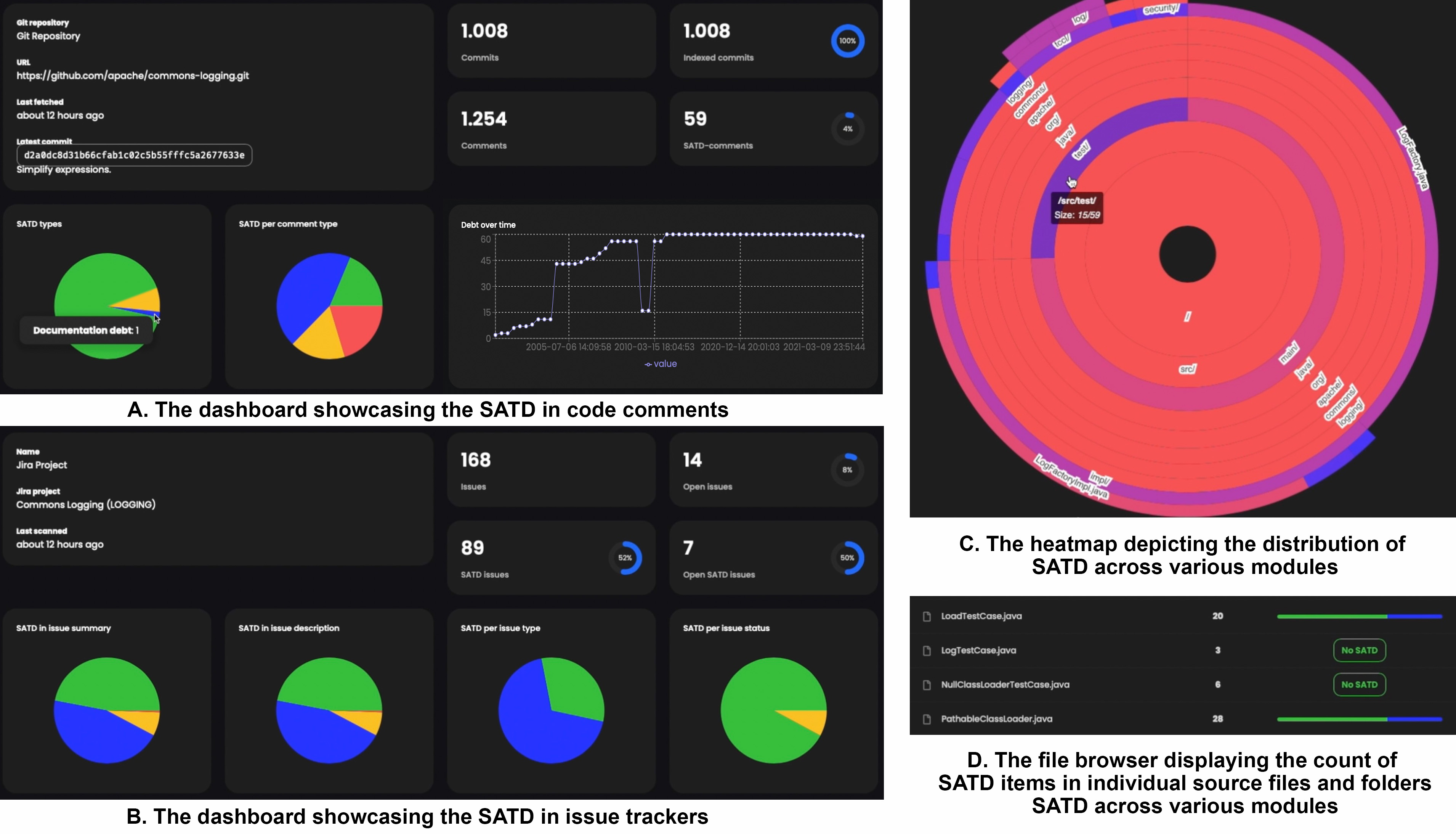}
\caption{The dashboard for SATD in code comments}
\label{f:screenshot}
\vspace{-5mm}
\end{figure*}

This paper introduces DebtViz, a SATD management tool capable of automatically detecting various types of SATD in source code comments and issue tracking systems using a Convolutional Neural Network-based approach.
In addition to identifying SATD, DebtViz generates and visualizes statistics on SATD items within software repositories.
Finally, the tool also serves as a real-time monitor for SATD: it constantly scans the software artifacts of a repository, updates the statistics, and visualizes the current state of SATD.
The four aforementioned functionalities map to four of the technical debt management activities as defined by Li \textit{et al.} \cite{li2015systematic}: TD \textit{identification}, \textit{measurement}, \textit{representation} and  \textit{monitoring}.

The architecture of DebtViz is comprised of three main components: a back-end service, an SATD classifier, and a front-end module.
The \textbf{back-end service} collects data from Git repositories and issue tracking systems, and subsequently pre-processes the data and populates it into a dedicated database.
The SATD classifier, the core part of the tool, employs a pre-trained machine learning model \cite{li2022automatic} to classify the types of SATD (i.e., code/design, test, documentation, and requirement debt), following the classifications suggested in prior work \cite{li2022automatic}.
Upon identification of SATD items, the classifier applies the deconvolution technique \cite{ren2019neural} to extract keywords that indicate the SATD classification.
The \textbf{front-end module} of DebtViz serves as the interface between the user and the tool, communicating with the back-end service to offer comprehensive SATD dashboards to the tool users.
It also provides a specialized browser for both issue trackers and source code.
DebtViz is designed for ease of deployment, scalability, and user-friendly interaction.
It supports the management of SATD, helping developers and managers gain a thorough understanding of SATD within their software systems.

The paper's structure is as follows: \cref{sec:related} discusses related work.
\cref{sec:overview} offers an expansive overview of our DebtViz tool, while
\cref{sec:architecture} outlines its architecture.
\cref{sec:evaluation} shows the results of a preliminary evaluation.
Conclusions and future directions are presented in \cref{sec:conclusion}.

\section{Related work}
\label{sec:related}

The first work exploring SATD was conducted by Potdar and Shihab \cite{potdar2014exploratory} in source code comments.
They analyzed four open-source projects and found that SATD comments were present in 2.4\% to 31\% of source files. 
Interestingly, they found that only a portion of SATD comments, ranging from 26.3\% to 63.5\%, were resolved after being documented.
Maldonado and Shihab \cite{maldonado2015detecting} extended this foundational work by refining the classification of SATD into five distinctive categories, namely design, requirement, defect, documentation, and test debt.
This classification was achieved by meticulously examining 33,000 code comments from five open-source projects.
Their results highlighted design debt as the most pervasive form of SATD, contributing to 42\% to 84\% of the categorized cases.

Subsequent to these initial explorations in the sphere of SATD, a considerable body of research has pivoted towards devising methods for automating SATD detection.
Several machine learning methodologies \cite{ren2019neural, li2022identifying, li2022automatic, guo2021far} have been utilized to detect diverse types of SATD instances from various sources.
Ren \textit{et al.} \cite{ren2019neural} proposed a Convolutional Neural Network-based method to improve SATD detection's accuracy and explainability, particularly enhancing cross-project prediction.
Similarly, Li \textit{et al.} \cite{li2022identifying} generated a dataset of 4,200 issues from seven open-source projects and proposed a machine learning approach to detect SATD in issue tracking systems, outperforming baseline methods, benefiting from knowledge transfer, and extracting intuitive SATD keywords.
Li \textit{et al.} \cite{li2022automatic} also proposed an automated SATD identification approach that leveraged a multitask learning technique to analyze multiple sources, including source code comments, commit messages, pull requests, and issue tracking systems.
Finally, Guo \textit{et al.} \cite{guo2019mat} introduced a straightforward heuristic approach for SATD identification, proving it to perform similarly or even better than existing methods.

Further research efforts have focused on creating tools to facilitate SATD management.
Liu \textit{et al.} \cite{liu2018satd} introduced a tool called SATD detector, capable of automatically detecting SATD comments using text mining techniques, and highlighting, listing, and managing detected comments in an integrated development environment.
This tool was designed with a back-end Java library and an Eclipse plug-in as its front-end.
Further expanding the toolbox for SATD management, Phaithoon \textit{et al.} \cite{phaithoon2021fixme} presented a GitHub bot specifically designed to handle issue-related \textit{On-hold SATD}, a situation in which developers delay proper implementation due to issues in the project issue tracker that are pending.
This bot leverages machine learning techniques to automatically detect On-hold SATD comments in source code and identify the referenced issues.
Upon resolution of the referenced issues, the bot notifies the developers accordingly.
In contrast to these earlier works, our research proposes a tool that: 1) focuses on identifying SATD from multiple sources, 2) offers a web-based dashboard to monitor SATD status in software systems, and 3) presents new visualizations of SATD data, such as heatmaps and SATD line charts.

\section{An overview of the DebtViz tool}
\label{sec:overview}

This section explains the main functionalities of the DebtViz tool through UI screenshots.
Utilizing the pre-trained SATD detection model \cite{li2022automatic}, DebtViz can automatically determine the types of SATD within code comments and issues in issue trackers.
Upon completion of data collection and analysis from a specific software repository, the tool generates two distinct dashboards: one presents details of SATD within code comments (subfigure A in \cref{f:screenshot}), while the other elaborates on SATD occurrences within issue tickets (subfigure B in \cref{f:screenshot}).

The dashboard depicted in subfigure A in \cref{f:screenshot} displays the variety of SATD types (i.e., code/design, test, requirement, and documentation debt) in code comments in the form of a pie chart, providing developers with a broad perspective of SATD categories prevalent in the software system. 
A secondary pie chart illustrates the distribution of SATD instances across different types of code comments, namely, inline, multi-line, block, and documentation block comments. 
Additionally, a line chart tracks the evolution of SATD in the system by representing the count of SATD items over time.

Similarly, the dashboard depicted in subfigure B in \cref{f:screenshot} serves to portray the variety of SATD categories prevalent within issue tracking systems.
As can be seen, the first two pie charts show the number of different types of SATD (i.e., code/design, test, requirement, and documentation debt) in the issue summary and issue description.
Moreover, it enumerates the instances of SATD occurring in different types of issues, such as tasks or bugs.
This detailed view allows developers to discern patterns of SATD accumulation within specific issue types, thereby aiding in strategic response planning.
Furthermore, the dashboard visualizes SATD items based on their status, for instance, distinguishing between open and closed issues.

To augment the understanding of SATD distribution across diverse modules, DebtViz generates comprehensive heatmaps, as exemplified in subfigure C in \cref{f:screenshot}. 
For example, we observe that the \textit{/src/test/} directory contains 15 SATD items out of a total of 59 detected within the entire system.
This functionality extends to various types of SATD, thereby offering developers and managers a visual representation of SATD dispersion throughout their software systems.

An additional key feature of DebtViz is a file browser functionality (subfigure D in \cref{f:screenshot}).
For example, we discern that \textit{LoadTestCase.java} contains SATD within a total of 20 comments, with code/design debt slightly outweighing test debt.
Conversely, there are no SATD items found within \textit{logTestCase.java}.
This enables a thorough enumeration of distinct SATD types across individual files and folders, thereby visualizing the distribution of varied SATD types across the software system.

\begin{figure}[htb]
\centering
\includegraphics[width=\linewidth]{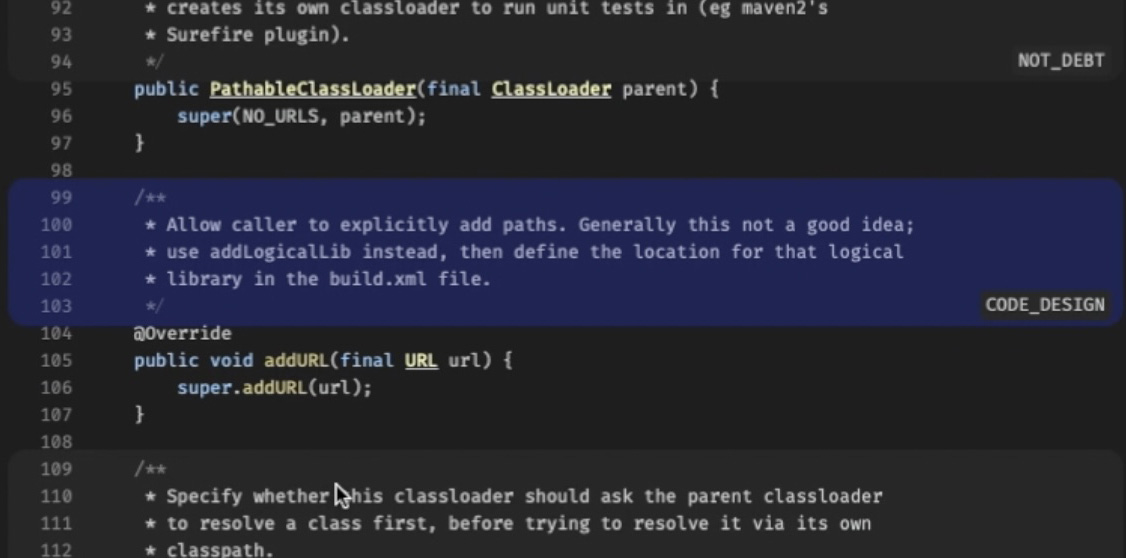}
\caption{The code viewer presenting the classification of different types of SATD or non-SATD for each code comment}
\label{f:browser-2}
\end{figure}

Upon selection of a specific source code file, DebtViz reveals the source code accompanied by a classification of SATD types for each code comment, as depicted in \cref{f:browser-2}. 
The tool distinctly highlights code comments classified as SATD, with the associated classifications such as code/design debt clearly marked. 
This feature facilitates developers in pinpointing the exact location of SATD within the source code.

\begin{figure}[htb]
\centering
\includegraphics[width=\linewidth]{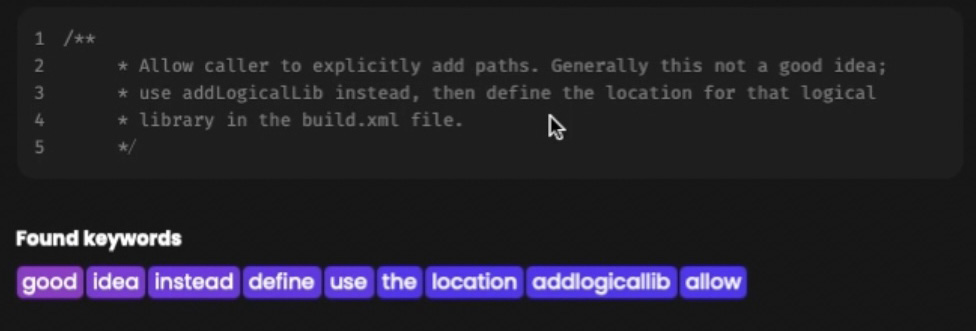}
\caption{Upon clicking the identified SATD code comments, the corresponding keywords highlighting the presence of SATD are displayed}
\label{f:keywords}
\end{figure}

\begin{figure*}[t]
\centering
\includegraphics[width=0.75\linewidth]{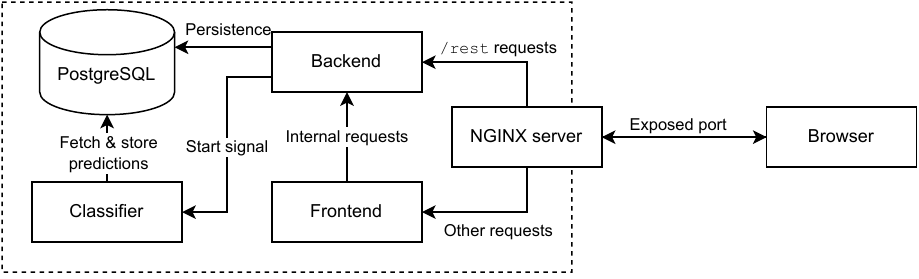}
\caption{The overview architecture of DebtViz}
\label{f:architecture}
\end{figure*}

Finally, DebtViz enables developers to dig deeper into the SATD code comments: when they select a code comment classified as SATD, DebtViz underscores the corresponding keywords that signal the presence of SATD, as demonstrated in \cref{f:keywords}.
By drawing attention to these keywords, developers are given insight into the potential root causes of the technical debt item.

\section{System architecture}
\label{sec:architecture}

The DebtViz tool architecture, as depicted in \cref{f:architecture}, comprises three primary components that will be elaborated in the following sub-sections: the back-end, the front-end, and the classifier.

\subsection{Back-end service}

This is the backbone of the DebtViz tool, and was developed utilizing the Spring framework.
This module is entrusted with scanning Git repositories for code comments and JIRA boards for issue tickets.
Upon extraction, the data is systematically stored in a PostgreSQL database.

The back-end service effectively leverages the JGit project, a Java implementation of the Git version control system, to store and manage files along with their associated revisions.
For the extraction of comments from source code files, we employed ANTLR (ANother Tool for Language Recognition).
ANTLR's grammar parsing capabilities facilitate the creation of simple grammars customized to the diverse types of comments we encounter in the code.

The process of scanning a JIRA project requires interfacing with the JIRA server’s REST API, a well-documented API with numerous open-source clients available.
In this project, we opted for the Jira REST Java Client (JRJC).

\subsection{Front-end service}

The front-end module of DebtViz is assigned with presenting an interactive and responsive user interface.
This module communicates with the back-end service, primarily by transmitting HTTP requests.
To create an engaging, high-performance UI, the front-end module was constructed using the versatile React library along with the Next.JS framework. The functionalities and UI of this module were discussed and presented more extensively in Section \ref{sec:overview}.

\subsection{SATD classifier}

The SATD classifier module scans the database for unclassified issues and comments, retrieving each entry for classification.
Using a pre-trained machine learning model \cite{li2022automatic}, it predicts the type of SATD (i.e., code/design, test, requirement, and documentation debt) for each unclassified entry.
Specifically, the pre-trained machine learning model leverages the multitask learning technique \cite{liu2015representation} in combination with Text-CNN \cite{kim2014convolutional}. 
This model is capable of detecting SATD from multiple sources, namely code comments, commit messages, issue trackers, and pull requests.
We note that DebtViz currently uses only code comments and issues, as these are the two most popular sources; commit messages and pull requests will be covered in the tool's next version.
In cases where the data is classified as SATD, the deconvolution technique \cite{ren2019neural} is employed to extract the likely causative keywords.
For instance, given a comment such as ``todo: we need to remove the dead code'', the extracted keywords could be ``todo'' and ``dead code''.
Once the extraction is complete, the prediction is saved back into the database.
The SATD classifier module incorporates a straightforward Flask server, for communication with the back-end service.
This setup ensures real-time data processing and classification, keeping SATD data up to date.

\section{Preliminary Evaluation}
\label{sec:evaluation}

\subsection{Evaluation setup}

In order to obtain some preliminary evidence on the usefulness of the DebtViz tool, we designed a small-scale study involving six developers.  
Specifically, we acquired their feedback on the use of DebtViz via a targeted survey (see the replication package \footnote{\url{https://github.com/yikun-li/visdom-satd-management-system}}).
The survey revolved around three main focal points: \textit{accuracy} pertains to the correctness of SATD detection by the tool; \textit{awareness} focuses on how well the tool helps users understand the state and distribution of SATD in the system; and \textit{effectiveness} assesses how much the tool assists users in managing SATD.
The questions were designed to not only provide individual feedback on each source (code comments or issue trackers) but also on using both sources.

We analyzed the results by first studying each focal point and then comparing the individual and combined scores of the two data sources (code comments and issue trackers).
This comparative analysis served to identify the efficacy of these data sources both independently and in tandem, offering insight into their individual strengths and the synergistic effects when used together.

\begin{figure}[htb]
\centering
\includegraphics[width=\linewidth]{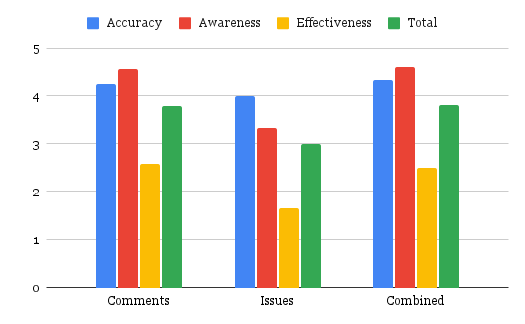}
\caption{Evaluation results}
\label{f:evaluation}
\vspace{-3mm}
\end{figure}

\subsection{Results}

In \cref{f:evaluation}, we provide a graphical representation of the average scores for each focal point. 
Regarding \textit{accuracy}, we found that the visualization of SATD in source code comments achieved a high score of 4.25 out of 5.
Participants reported only a few instances of inaccurate classification, primarily attributed to misleading keywords in comments. 
Meanwhile, the visualization of JIRA issues garnered a good average score of 4 out of 5.
Participants reported occasional discrepancies between the classification of the summary and description, once again tracing the errors back to certain keywords.
The combined accuracy, however, was rated even higher at 4.33 out of 5, reflecting the tool's robust capacity to detect SATD in software projects.

Turning our attention to the \textit{awareness} generated by the tool, participants noted that they discovered forgotten areas of interest within source code comments.
The SATD visualization in comments scored an impressive average of 4.55 out of 5, with participants recognizing the potential value of tracking SATD over a longer period. 
Visualization of SATD in issues received a lower score of 3.33, as participants felt that issues were already well-sorted and organized within JIRA.
Overall, the tool scored highest in enhancing awareness, reaching 4.61 out of 5.

The final focus point, \textit{effectiveness}, received a lower rating from participants, with an average score of 2.58 out of 5 for comments and 1.66 out of 5 for issues, culminating in a combined average of 2.5.
The participants argued that the visualized SATD had already existed in their work for a longer time, thus it did not have a high priority for refactoring.
They also stated that they already prioritized SATD items using their issue tracker, so using our tool would not effectively assist in this task.

Finally, when evaluating the combined total rating of both sources, the average score rose to 3.81, indicating a slight advantage when utilizing both sources concurrently, compared to the separate scores of 3.80 and 3.00 for the comments and issues respectively.

\section{Conclusion}
\label{sec:conclusion}

This paper presented DebtViz, a tool designed specifically to detect, classify, monitor and visualize various types of SATD in source code comments and issue tracking systems.
The system comprises a back-end service, a SATD classifier, and a front-end module for interactive visual exploration. 
DebtViz aids in the management of SATD by providing an overview of the current state of SATD within software systems, which in turn supports informed decision-making regarding technical debt management.
Potential areas for future enhancement include expanding the range of software artifacts (e.g., pull requests and commit messages) considered in SATD detection and refining the machine learning models for improved classification accuracy.
As a contribution to the field of SATD management, DebtViz offers an integrated, multifaceted approach to identifying and visualizing SATD across multiple sources.

\bibliography{main}
\bibliographystyle{IEEEtran}

\end{document}